\documentclass[preprint,superscriptaddress,amsmath,amssymb,longbibliography]{revtex4-1}
\usepackage{graphicx}
\usepackage{bm}
\usepackage[colorlinks=true,urlcolor=blue,citecolor=blue,linkcolor=blue,bookmarks=false,pdfstartview={FitH}]{hyperref}
\usepackage{color}

\newcommand{\be}{\begin{equation}}
\newcommand{\ee}{\end{equation}}

\newcommand{\bea}{\begin{eqnarray}}
\newcommand{\eea}{\end{eqnarray}}

\renewcommand{\vec}[1]{{\bf #1}}

\definecolor{mycyan}{rgb}{0.0, 0.5, 0.5}

\def \cm-1{cm$^{-1}$}

\begin{document}
\title{
Polar charge density wave in a superconductor with crystallographic chirality}
\author{Shangfei~Wu}
\email{wusf@baqis.ac.cn}
\thanks{Present address: Beijing Academy of Quantum Information Sciences, Beijing 100193, China.}
\affiliation{Department of Physics and Astronomy, Rutgers University, Piscataway, New Jersey 08854, USA}
\author{Fei-Ting~Huang}
\affiliation{Department of Physics and Astronomy, Rutgers University, Piscataway, New Jersey 08854, USA}
\affiliation{Keck Center for Quantum Magnetism, Rutgers University, Piscataway, New Jersey 08854, USA}
\author{Xianghan~Xu}
\thanks{Present address: School of Physics and Astronomy, University of Minnesota, Minneapolis, MN, USA.}
\affiliation{Department of Physics and Astronomy, Rutgers University, Piscataway, New Jersey 08854, USA}
\affiliation{Keck Center for Quantum Magnetism, Rutgers University, Piscataway, New Jersey 08854, USA}
\author{Ethan T.~Ritz} 
\thanks{Present address: Department of Engineering, Harvey Mudd College, Claremont, CA, USA}
\affiliation{Department of Chemical Engineering and Materials Science, University of Minnesota, Minneapolis, MN 55455, USA}
\author{Turan~Birol} 
\email{tbirol@umn.edu}
\affiliation{Department of Chemical Engineering and Materials Science, University of Minnesota, Minneapolis, MN 55455, USA}
\author{Sang-Wook~Cheong} 
\email{sangc@physics.rutgers.edu}
\affiliation{Department of Physics and Astronomy, Rutgers University, Piscataway, New Jersey 08854, USA}
\affiliation{Keck Center for Quantum Magnetism, Rutgers University, Piscataway, New Jersey 08854, USA}
\author{Girsh~Blumberg} 
\email{girsh@physics.rutgers.edu}
\affiliation{Department of Physics and Astronomy, Rutgers University,
Piscataway, New Jersey 08854, USA}
\affiliation{National Institute of Chemical Physics and Biophysics, 12618 Tallinn, Estonia}

\date{\today}
\maketitle

\textbf{
Symmetry plays an important role in determining the physical properties in condensed matter physics, as the symmetry operations of any physical property must include the symmetry operations of the point group of the crystal. As a consequence, crystallographic polarity and chirality are expected to have an impact on the Cooper pairing in a superconductor. While superconductivity with crystallographic polarity and chirality have both been found in a few crystalline phases separately; however, their coexistence and material realizations have not been studied. Here, by utilizing transport, Raman scattering, and transmission electron microscopy, we unveil a unique realization of superconductivity in single-crystalline Mo$_3$Al$_2$C (superconducting $T_\text{c}$=8\,K) with a polar charge-density-wave phase and well-defined crystallographic chirality. We show that the intriguing charge density wave order leads to a noncentrosymmetric-nonpolar to polar transition below $T^*$=155K via breaking both the translational and rotational symmetries. Superconductivity emerges in this polar and chiral crystal structure below $T_\text{c}$=8\,K. Our results establish that Mo$_3$Al$_2$C is a superconductor with crystallographic polarity and chirality simultaneously, and motivate future studies of unconventional superconductivity in this category.
}

The concept of `ferroelectric metals' was first proposed by Anderson and Blount more than half a century ago~\cite{Anderson_1965_PhysRevLett}. Ferroelectricity and metallicity are conventionally considered incompatible as itinerant electrons tend to screen long-range Coulomb interactions and weaken polar instabilities. This makes the appearance of an inversion-symmetry breaking transition in metals rare, although numerous compounds are predicted to host such transitions~\cite{Benedek_2016_JMCC,Zhou_2020_review,Ghosez_2022_review,Bhowal_arxiv2022_review,Young_2023PRM, Li2021}. Despite these challenges, LiOsO$_3$ and WTe$_2$/MoTe$_2$ are two unambiguous examples of experimentally verified polar metals, driven by polar instability which is decoupled with itinerant electrons at the Fermi level~\cite{Shi_2013_NatMat,Puggioni_2014_NC,Kim_2016_nature} and in-plane interlayer sliding mechanisms~\cite{Fei_2018_nature,MoTe}, respectively. 
Another way to generate polarization is via translational symmetry breaking by virtue of a charge-density wave (CDW). 
When a non-centrosymmetric charge modulation occurs, it could result in a potential net electric polarization~\cite{Cheong_2007_review,van_den_Brink_2008_review,Fiebig_2016_review}. 
Achieving CDW-driven polar metals have particular significance that they can in principle 
enable ultrafast switching at characteristic electronic rather than lattice time scales~\cite{Qi_2022PRB,Liu_arxiv2023}.
                                                                                                                                                            
Since superconductivity (SC) tends to arise from itinerant electrons in a metallic system, the coexistence of ferroelectricity/polarity and superconductivity is unusual and is a topic of active discussions~\cite{Enderlein_2020NC,Salmani_Rezaie2020,Hameed_2022NatureMaterials,Jindal_2023Nature}. Examples of such polar superconductors (existence of a polar axis in a superconductor) are the doped SrTiO$_3$~($T_\text{c}$ less than 1\,K)~\cite{Enderlein_2020NC,Salmani_Rezaie2020,Hameed_2022NatureMaterials}, bilayer T$_\text{d}$-MoTe$_2$ ($T_\text{c}$=2.3\,K)~\cite{Jindal_2023Nature}, and various heavy fermions superconductors ($T_\text{c} \sim$1-3\,K)~\cite{Smidman_2017}. The presence of a polar axis in a superconductor gives rise to phenomena such as exotic nonreciprocal charge transport~\cite{Edelstein_1996,Tokura_2018NC,Yuki_2020sciadv,Nagaosa_2024review}, including the superconducting diode effect~\cite{Nadeem_2023review},  the mixing of spin-triplet and spin-singlet order parameters, as well as topological superconductivity featuring a Majorana edge state~\cite{Yip_2014Review}. On the other hand, superconductors with a chiral structure are reported in Li$_2$Pd$_3$B, Li$_2$Pt$_3$B~\cite{Yuan_2006PhysRevLett,Togano_2004PhysRevLett}, Mo$_3$Al$_2$C~\cite{Karki_2010_PhysRevB}, NbRh$_2$B$_2$~\cite{Elizabeth_2018Science_Advances}, and TaRh$_2$B$_2$~\cite{Elizabeth_2018Science_Advances}. Among them,  Li$_2$Pt$_3$B has been proposed to be a candidate for chiral superconductivity~\cite{Yuan_2006PhysRevLett}.
                  


Superconductors with a polar and chiral crystal structure are expected to have unconventional Cooper paring, as the superconducting order parameter must respect the symmetry of the underlying crystal lattice according to the Neumann's principle. However, such type of superconductors is rare.
Identifying superconductors in this category are important not only for fundamental science in the multiferroic and superconductivity community, but also for technological applications~\cite{Nadeem_2023review}.

In this work, we present evidence for a new category of superconductor, Mo$_3$Al$_2$C, which exhibits superconducting transition temperature ($T_\text{c}$) of 8\,K and possesses both structural polarity and chirality. The polarization is induced by a CDW order that appears below $T^*$ = 155\,K, characterized by a cubic-nonpolar to rhombohedral-polar phase transition. Superconductivity emerges deep within this polar and chiral crystal structure below $T_\text{c}$, establishing Mo$_3$Al$_2$C as a unique superconductor with both structural polarity and chirality.

\begin{figure*}[!t] 
\begin{center}
\includegraphics[width=\columnwidth]{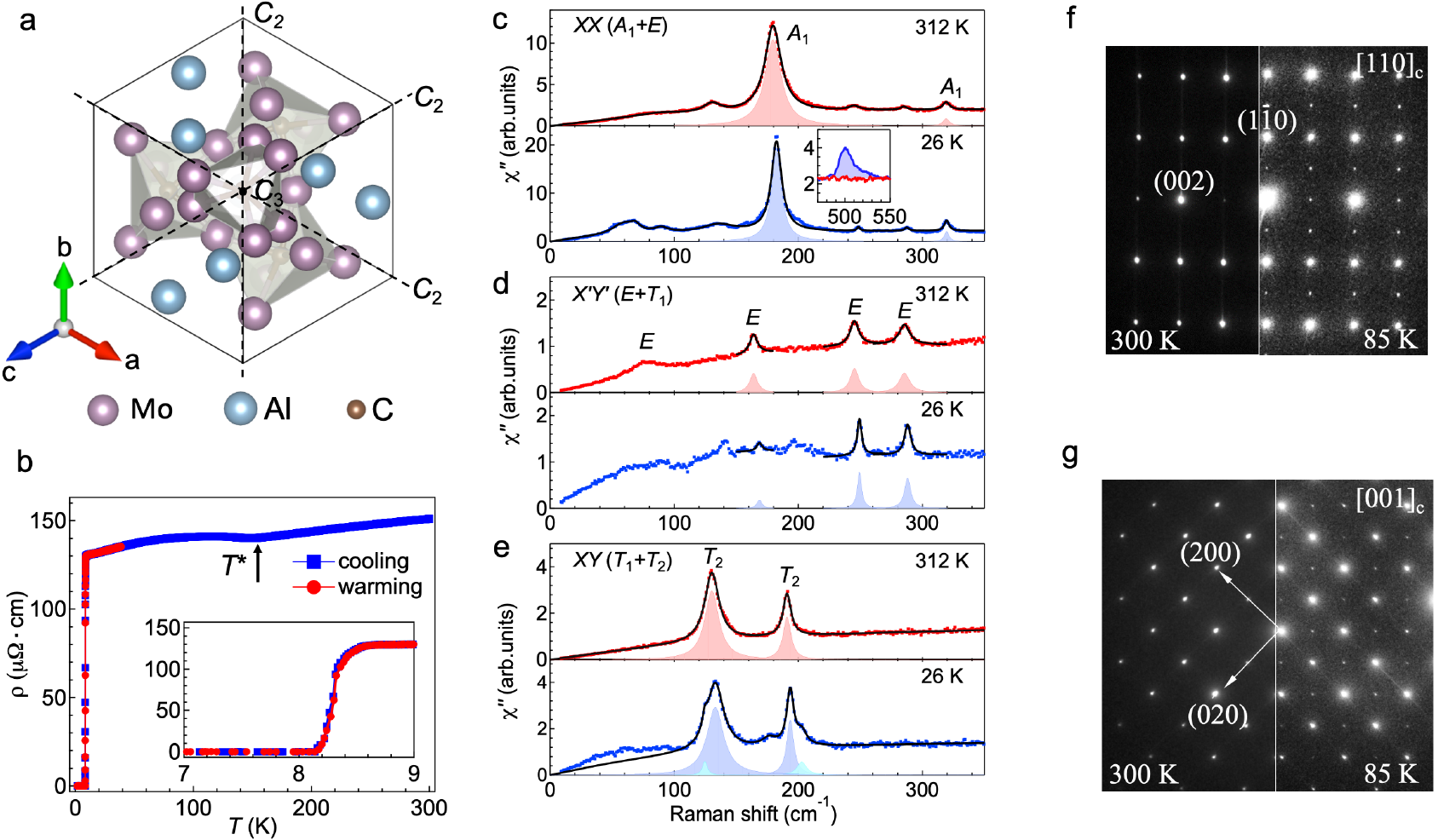}
\end{center}
\caption{\label{Fig1_structure} \textbf{Crystal structure of Mo$_3$Al$_2$C}. \textbf{a} Crystal structure of Mo$_3$Al$_2$C showing one $C_3$ axis and three $C_2$ axis perpendicular to the three-fold axis. \textbf{b} Resistivity of Mo$_3$Al$_2$C on the (1~0~0) surface recorded in the cooling-down process (blue squares) and warming-up process (red circles). The inset of (b) shows the resistivity around superconducting transition temperature. 
\textbf{c,d,e} Raman spectra of Mo$_3$Al$_2$C on a polished (1~0~0) surface in the $XX$ (c), $X'Y'$ (d), and $XY$ (e) scattering geometries at 312\,K and 26\,K. The black solid lines as well as the red and blue shaded areas are the fitting results.
\textbf{f,g} Temperature dependent selected area electron diffraction patterns of Mo$_3$Al$_2$C single crystal along [1~1~0]$_\text{c}$/[1~$\bar{1}$~$\bar{4}$]$_\text{h}$ (f)  and [0~0~1]$_\text{c}$/[1~$\bar{1}$~2]$_\text{h}$ (g), revealing superlattice peaks below $T^*$. The subscript $\text{c}$ and $\text{h}$ denote the cubic and hexagonal notations, respectively.
}
\end{figure*} 
 
\textbf{Results}
   
\textbf{Crystal structure.}
Mo$_3$Al$_2$C has a cubic structure with space group $P4_132$ or its enantiomorphic space group $P4_332$ (point group
$O$) at room temperature (Supplementary Note~I). The crystal structure lacks inversion and mirror symmetries. There are four three-fold axes along the body-diagonal directions and three $C_2$ axes perpendicular to a fixed three-fold axis as shown in Fig.~\ref{Fig1_structure}a. Thus, the structure of Mo$_3$Al$_2$C is noncentrosymmetric, chiral, and nonpolar at room temperature. Upon cooling, as shown in Fig.~\ref{Fig1_structure}b, resistivity measurement shows that there is a superconducting transition at around $T_\text{c}$=8\,K and bulk superconductivity is confirmed by the magnetic susceptibility measurement (Supplementary Note~I). The resistivity measurement
also shows a dip at around $T^*$=155\,K~\cite{Zhigadlo_2018PRB}. The anomalies at $T^*$ are also found in earlier magnetic susceptibility, specific heat, nuclear-magnetic-resonance (NMR) measurements~\cite{Koyamado_2013_JPSJ,Koyama2011PhysRevB}.

\begin{figure*}[t] 
\begin{center}
\includegraphics[width=\columnwidth]{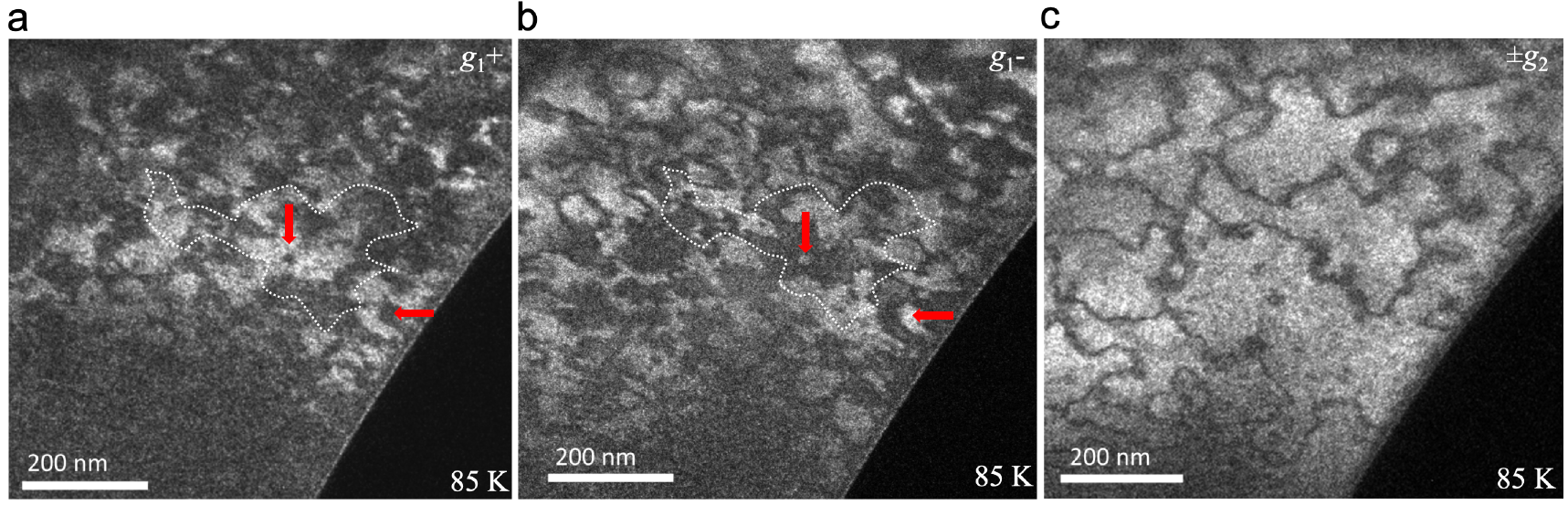}
\end{center}
\caption{\label{Fig2_polar_domain} 
\textbf{Polar domain and anti-phase domain}. \textbf{a} Dark-field TEM images were taken at 85\,K using $g_1^+= (1/2, -1/2, 4)_\text{c}$ spot along [1~1~0]$_c$.  Abundant 180$^{\circ}$ polar domains on the 100\,nm scale or less are shown in black and white contrast. Scale bar is 200\,nm.
\textbf{b} Dark-field-images acquired at the same area using the $g_1^-= -(1/2, -1/2, 4)_\text{c}$ spot.
\textbf{c} Dark-field-images acquired at the same area using the $g_2^{\pm}=\pm(1/2, -1/2, 0)_\text{c}$ spots. Red arrows and a white dot loop are added to guide the eye. The reversed contrast between (a) and (b) demonstrates the Friedel's pair breaking due to the noncentro-symmetrical polar nature, while no contrast change of antiphase domains shown in (c). The average polar domains are on the order of 100\,nm while antiphase domains are about half-micrometer.
}
\end{figure*}  
                                                                                                                                                                                                                                                                                                   
To understand the resistivity anomaly at $T^*$, we perform Raman scattering measurement from a polished (1~0~0) surface of Mo$_3$Al$_2$C in the $XX$, $X'Y'$, and $XY$ scattering geometries at 312\,K and 26\,K. For the $XX$ scattering geometry shown in Fig.~\ref{Fig1_structure}c, we detect two phonon modes at 180 and 319\,\cm-1 with Lorentzian lineshape at 312\,K and 26\,K, and a third phonon mode at around 500\,\cm-1 that appears only at low temperature.
These three $A_1$-symmetry modes are dominated by Mo, Al and C lattice vibrations, respectively~\cite{Reith_2012PhysRevB} (Supplementary Note~II). 
Specifically, the broad peak at around 50\,\cm-1 appearing at low temperatures is the amplitude mode of the CDW order parameter (see the section of the CDW transition). The peak at 129\,\cm-1 at 312\,K in the $XX$ scattering geometry is a leakage of the $T_2$ phonon from the $XY$ scattering geometry due to imperfect cutting or sample polishing of the single crystal.
In the $X'Y'$ scattering geometry, we detect three sharp phonon modes at 164, 245, and 285\,\cm-1 and a broad mode at 75\,\cm-1 at 312\,K shown in Fig.~\ref{Fig1_structure}d. They are all $E$-symmetry phonon modes. 
The three sharp $E$ modes are first-order phonons, while the broad one is a second-order two-phonon excitation as the linewidth is four times of the three sharp ones.
Below $T^*$, the three first-order $E$-symmetry phonon modes do not split at 26\,K. 
In the $XY$ scattering geometry, we detect two $T_2$-symmetry phonons at 129 and 190\,\cm-1 at 312\,K. They split into two modes in the 26\,K data as shown in Fig.~\ref{Fig1_structure}e, indicating a structural phase transition below $T^*$.      
Since the $T_2$ phonon splitting is independent of the domain orientations, the presence of domains does not affect the conclusion.                       
By virtue of the correlation table of the point group $O$ of the high-temperature structure~(METHODS), based on the fact that $E$ modes do not split while $T_2$ modes split into two modes below $T^*$, we infer that low-temperature structure belongs to a rhombohedral point group $C_3$ or $D_3$. 
                                                                                                                                                                   
To reveal the potential ordering vector of the low-temperature structure, we conduct transmission electron microscopy (TEM) measurement on Mo$_3$Al$_2$C.
Fig.~\ref{Fig1_structure}f-g represent the [1~1~0]$_\text{c}$ and [0~0~1]$_\text{c}$ zone-axis selected area electron diffraction patterns taken at 300\,K and 85\,K, respectively. The data at 300\,K [left panels of Fig.~\ref{Fig1_structure}f-g] are consistent with the primitive $P4_132$ cubic structure ($a= 6.8638$\,\AA). 
The data at 85\,K, shown in the right panels of Fig.~\ref{Fig1_structure}f-g, reveal evidence of a phase transition from the presence of the systematic superlattice Bragg peaks alongside the fundamental Bragg peaks. 
The appearance of the superlattice reflections (1/2, -1/2, 0) [Fig.~\ref{Fig1_structure}f] and (1/2, -1/2, 0)/(-1/2, -1/2, 0) [Fig.~\ref{Fig1_structure}g] while they are absent at half of (0, 0, 1) indicates a transition involving $2\times2\times1$ supercell in the hexagonal setting. Our results suggest that Mo$_3$Al$_2$C undergoes a transition to a rhombohedral structure below $T^*$, leading to either the nonpolar R32 (point group $D_3$) or the polar R3 (point group $C_3$) space group with lattice vectors (-2, 2, 0), (0, -2, 2), and (1, 1, 1) in relation to those of the cubic unit cell.
The lattice constants for the supercell are $a \approx 11.896$\,\AA, $\alpha=109.58 ^{\circ}$ in the trigonal setting or $a \approx 19.44$\,\AA, $c \approx 11.83$\AA~in the hexagonal setting. 

\begin{figure*}[t] 
\begin{center}
\includegraphics[width=\columnwidth]{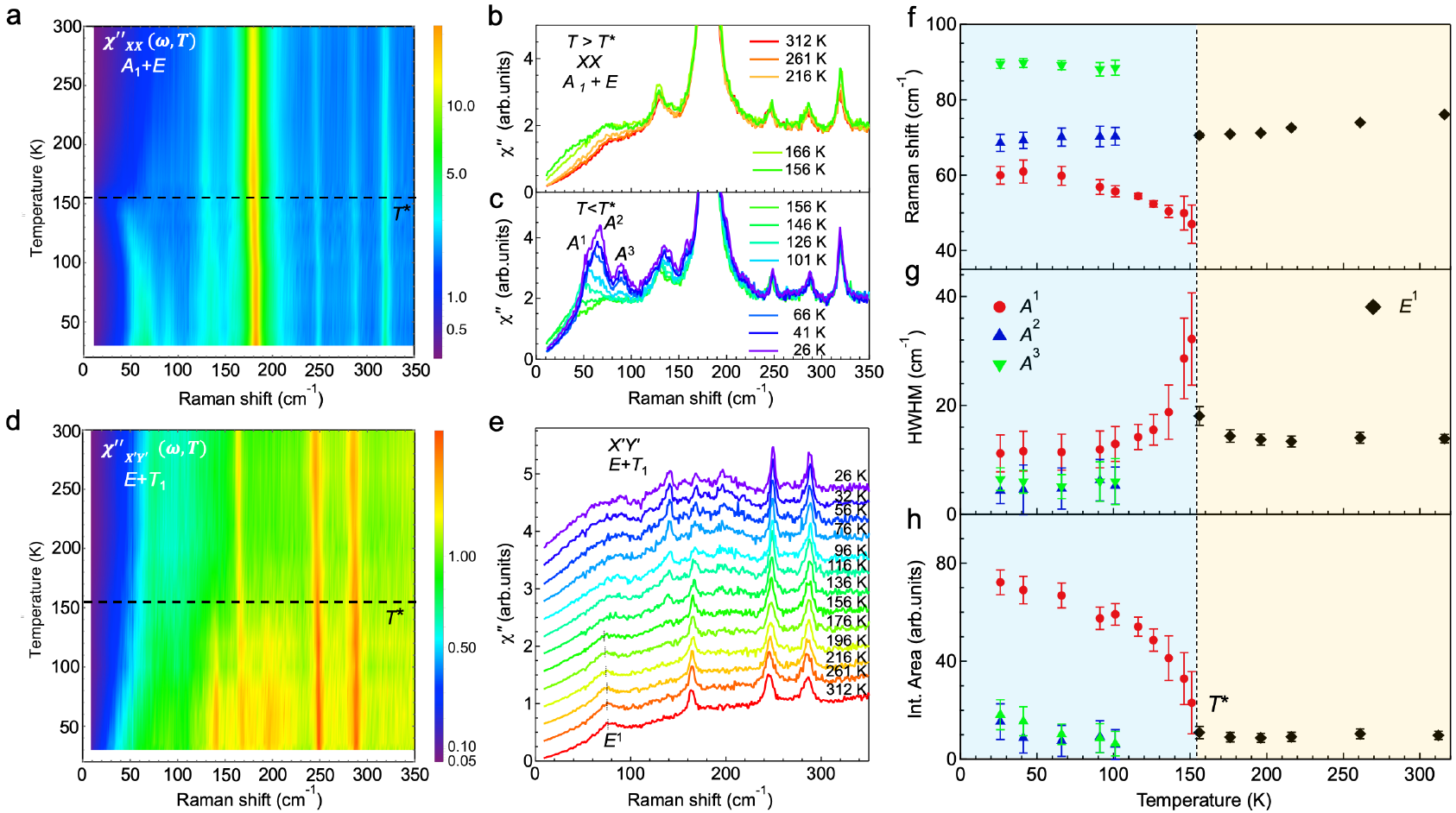}
\end{center}
\caption{\label{Fig3_CDW} \textbf{Raman data recorded on a polished (1~0~0) surface of Mo$_3$Al$_2$C}.
\textbf{a} The colorplot of the $T$-dependence of Raman spectra in $XX$ [$A_1+E$] scattering geometry. 
\textbf{b} $T$-dependence of Raman spectra for $T>T^*$ corresponding to (a).
\textbf{c} Same as (b) but for $T<T^*$.
\textbf{d} The colorplot of the $T$-dependence of Raman spectra in $X'Y'$ [$E+T_1$] scattering geometry. 
\textbf{e} $T$-dependence of Raman spectra corresponding to (d). Spectra are vertically shifted for better visualizations. 
\textbf{f,g,h} $T$-dependence of peak frequency, half-width-at-half maximum (HWHM), and integrated area for the $A^1$, $A^2$, $A^3$ and $E^1$ modes marked in (c) and (e).
Error bars represent the standard deviation.
}
\end{figure*}

\textbf{Polar domains.}          
An intriguing question arises: Is the superconductor Mo$_3$Al$_2$C with structural chirality also a polar superconductor? 
To address this question and probe polar domains in Mo$_3$Al$_2$C, we utilize dark-field transmission electron microscopy (DF-TEM), which is known for its high spatial resolution and ability to isolate distinct types of domains using a specific diffraction spots~\cite{Huang_2019NC}. In Fig.~\ref{Fig2_polar_domain}, we show a series of DF-TEM images taken along the [1~1~0]$_\text{c}$ or [1~$\bar{1}$~$\bar{4}$]$_{\text{h}}$ direction using superlattice peaks (1) $g_1^{\pm}=\pm(1/2, -1/2, 4)_\text{c}$ containing the polar contributions along the [1~1~1]$_\text{c}$ polar axis [Fig.~\ref{Fig2_polar_domain}a-b], and (2) $g_2^{\pm}=\pm(1/2, -1/2, 0)_\text{c}$ containing mainly the information in the plane perpendicular to the [1~1~1]$_\text{c}$ polar axis [Fig.~\ref{Fig2_polar_domain}c]. 
Evident in Fig.~\ref{Fig2_polar_domain}a-b is the presence of 180$^{\circ}$-type polar domains on a 100-nm scale, showing broken fourfold and two-fold symmetry of the cubic phase, resulting in the polar R3 phase with point group $C_3$. 
The polar domains with black and white contrasts are associated with the $\pm$[1~1~1]$_\text{c}$ or $\pm[0~0~1]_\text{h}$ polar axes, approximately 35$^{\circ}$ away from the out-of-plane direction of the images. In addition to the small polar domains, sub-micrometer-sized wavey antiphase boundaries shown as dark lines are resolved [Fig.~\ref{Fig2_polar_domain}c]. 
These antiphase boundaries do not exist above $T^*$ and exhibit varying features during different cooling cycles, suggesting that domain formation below $T^*$  is not simply due to pinning by disorder such as chemical defects or dislocations (Supplementary Note~III). 
These images show that there seems to be no clear interlockings for these two types of domains below $T^*$. 
Our results reveal a non-polar to polar transition below $T^*$ in the superconductor Mo$_3$Al$_2$C. 

\textbf{CDW transition.} 
To further uncover the nature of phase transition at $T^*$, we study the temperature dependence of the Raman response in the $XX$ and $X'Y'$ scattering geometries.         
In Fig.~\ref{Fig3_CDW}a-b, we show the Raman response in the $XX$ scattering geometry. Upon cooling to $T^*$, the low-energy electronic continuum below 100\,\cm-1 builds up gradually in the $XX$ ($A_1+E$) scattering geometry while the continuum remains unchanged for the $X'Y'$ ($E+T_1$) scattering geometry~(Supplementary Note~IV), indicating the existence of the $A_1$-type fully-symmetric charge fluctuations. This is a signature of a precursor density-wave order approaching $T^*$, similar to the CDW-compound ErTe$_3$~\cite{Eiter2013PNAS}.  
Below $T^*$ as shown in Fig.~\ref{Fig3_CDW}c, the most prominent feature is the redistribution of the low-energy spectra weight: The low-energy continuum intensity below 40\,\cm-1 is progressively suppressed, while the high-energy continuum intensity in the range between 40\,\cm-1 and 250\,\cm-1 is enhanced and a broad peak centered at around 60\,\cm-1 (labeled as $A^1$) is developed. Such a spectra weight transfer is an evidence for gap opening below $T^*$~\cite{Devereaux2007RMP}.  
This observation is consistent with gap opening from the previous $^{27}$Al NMR studies that the Knight shift and the spin-lattice relaxation rates are significantly reduced below $T^*$~\cite{Koyama2011PhysRevB,Kuo_2012PhysRevB}.  
Therefore, the existence of charge fluctuations above $T^*$, the gap opening below $T^*$, and superlattice Bragg peaks appearance below $T^*$ all point to a second-order CDW transition at $T^*$. Moreover, as the temperature cools below 100\,K, two shoulder peaks at around 68 (labeled as $A^2$) and 90\,\cm-1 (labeled as $A^3$) become noticeable and gain intensity on top of the $A^1$ peak. 
                                                                       
In Fig.~\ref{Fig3_CDW}d-e, we show the Raman response in the $X'Y'$($E+T_1$) scattering geometry. In addition to the three fundamental $E$ modes at 164, 245, and 285\,\cm-1 which persists upon cooling from 312\,K to 26\,K, several new $E$-symmetry modes between 140\,\cm-1 and 220\,\cm-1 appear below $T^*$.                                                                                                                                                                                                                                                                                                                                                                                                                                                                                                                                                                                         
Notably, the two-phonon excitation feature (labeled as $E^1$) softens a bit upon cooling to $T^*$.

In Fig.~\ref{Fig3_CDW}f-h, we present the temperature dependence of the mode energy, half-width-at-half maximum (HWHM), and the integrated area for the $A^1$, $A^2$, $A^3$, and $E^1$ modes.
The $A^1$ mode frequency increases from 48 to 60\,\cm-1 while $A^2$ and $A^3$ modes barely change upon cooling below $T^*$. In contrast, the $E^1$ mode softens from 75\,\cm-1 at 312\,K to 70\,\cm-1 at $T^*$.
The HWHM for the $A^1$ mode decreases by a factor of three, while $A^2$ and $A^3$ modes barely change upon cooling. The integrated area for the $A^1$ mode increases gradually below $T^*$, reaching four times of its value at $T^*$ when cooled to the lowest temperature, while the $A^2$ and $A^3$ modes' integrated area increase moderately below $T^*$.
The differences in the temperature dependence between $A^1$, $A^2$, and $A^3$ peaks suggest that the $A^1$ peak represents the CDW amplitude mode while $A^2$ and $A^3$ peaks correspond to the folded phonon modes~\cite{mialitsin2010raman,Snow_2003PhysRevLett}.

Upon cooling below $T_\text{c}$, the phononic features barely change, namely, the amplitude mode $A^1$ and folded phonon modes $A^2$ and $A^3$ remain the same~(Supplementary Note~V). This indicates that the CDW order coexists with superconductivity. Furthermore, since Raman scattering is exceptionally sensitive to the subtle breaking of crystalline symmetry~\cite{Devereaux2007RMP,Wu_2024PRX}, the persistence of the same phononic features above and below $T_\text{c}$ indicates that the point group symmetry ($C_3$) persists also in the SC state. Thus, the SC state of Mo$_3$Al$_2$C is both chiral and polar.

\textbf{Discussions}

\textbf{Origin of the CDW transition.}
We now discuss the origin of the CDW transition at $T^*$ in Mo$_3$Al$_2$C. 
Based on the density functional theory (DFT) calculations~(Supplementary Note~VI), there are three unstable modes in the phonon dispersion for the cubic phase of Mo$_3$Al$_2$C: $\Gamma_4$(0,~0,~0), $M_5$(0.5,~0.5,~0), and $X_2$(0.5,~0.5,~0), consistent with Ref.~\cite{Reith_2012PhysRevB}. 
The $\Gamma_4$ instability leads to a R3 structure without translational-symmetry breaking, and hence the same unit-cell volume. The only structure with 3-fold rotational symmetry that the $X_2$ instability can lead to is a $2 \times 2 \times2$ supercell with 8 times the volume of the primitive cell~(Supplementary Note~VII). 
On the other hand, the $M_5$ instability can lead to the R3 structure with the correct translational symmetry breaking and 4 times the volume of the primitive cell.
Thus, the only way to get a CDW structure below $T^*$ commensurate with  $2\times2\times1$ supercell in the hexagonal setting and 3-fold rotational symmetry in the R3 phase is to have a $M_5$ lattice instability at the Brillouin zone boundary.
The observed softening of the second-order $E^1$ mode above $T^*$ [Fig.~\ref{Fig3_CDW}f] suggests the softening phonon branch at around $M$ points, similar to the study of 2H-NbSe$_2$~\cite{mialitsin2010raman}.

\textbf{Origin of the polarization.}
The origin of the polarization in Mo$_3$Al$_2$C poses new questions. Since the crystal structure is already noncentrosymmetric above $T^*$, the transition to the polar structure is not a simple off-centering of atoms, but rather roots in the breaking of the two-fold rotational symmetries perpendicular to the polar axis in the symmetry group of the re-arranged atoms below $T^*$~[Fig.~\ref{Fig1_structure}a]. It can nevertheless be either due to a zone-center instability that breaks the symmetry (proper ferroelectricity), or a zone-boundary instability that breaks translational symmetry in addition to the rotational symmetry (improper ferroelectricity). In the former case, another important question arises about whether the transition is a displacive or an order-disorder type transition. 
         
The order-disorder mechanism was discussed for the case of LiOsO$_3$~\cite{Liu_2015PhysRevB,Hyunsu_2014PhysRevB}. One of the key evidence in support of this mechanism in LiOsO$_3$ is the sharp drop in resistivity below the polar transition. The large resistivity above the transition is attributed to the incoherent charge transport induced by the disordered Li off-center displacements~\cite{Shi_2013_NatMat}. For Mo$_3$Al$_2$C, this mechanism for the nonpolar-to-polar transition is less likely, because the temperature dependence of the resistivity is almost flat from 300\,K down to 8\,K for single crystals [Fig.~\ref{Fig1_structure}b]~\cite{Zhigadlo_2018PRB} as well as polycrystalline samples~\cite{Bauer_2010_PhysRevB}. While some amount of disorders or vacancies could be present in the Mo$_3$Al$_2$C sample~\cite{Zhigadlo_2018PRB}, 
the flat resistivity curve suggests that the scattering rates for the electrons barely change upon cooling through the transition, which rules out disorders near the transition, and hence an order-disorder mechanism.
                                             
Our DFT lattice dynamics calculations, as well as earlier report~\cite{Reith_2012PhysRevB}, indicate that there are both zone-center and zone-boundary instabilities, which makes both proper and improper instabilities possible. The unstable $\Gamma$ mode transforms as the $\Gamma_4$ irreducible representation. It is infrared-active and a polar instability. Thus, the zone-center $\Gamma_4$ mode leading to the polarization while a zone-boundary $M_5$ mode leading to the CDW order is seemingly possible. However, this scenario is inconsistent with the observation that there is only a single transition at $T^*$.  It is also not possible for the CDW order being a secondary and improper order parameter that condenses only due to the nonzero amplitude of the polar mode, because the CDW order breaks the translational symmetry, which by definition cannot be broken by a zone-center $\Gamma$ polar mode.

This may lead us to conclude that the polarization is induced by an improper mechanism, where a zone-boundary $M_5$ mode couples with a zone-center  $\Gamma_4$ polar mode, and leads to a nonzero magnitude of the polarization below $T^*$. The details of this coupling can be understood by building a Landau free-energy expansion for the high-temperature cubic structure via the trilinear coupling between the $M_5$ and the polarization $\vec{P}$ along the [1~1~1] pseudo-cubic direction~(Supplementary Note~VIII).  
When the $M_5$ mode condenses, this leads to a first-order term in $\vec{P}$ to appear in the free energy expansion, which necessarily leads to a nonzero value of $\vec{P}$~(Supplementary Note~VIII).

We note that the density of states at the Fermi level are dominated by Mo $4d$ electronic states from the DFT band structure calculations~\cite{Karki_2010_PhysRevB}. In contrast, the Al and C electronic states are negligibly small between -2\,eV and 2\,eV, and they get a bit larger below -2\,eV~\cite{Karki_2010_PhysRevB}. Furthermore, from the DFT phonon calculations~\cite{Reith_2012PhysRevB}, the phonon density of states below 100\,\cm-1 are dominated by Mo lattice vibration modes, while they are dominated by C lattice vibration modes above 500\,\cm-1. In between 100\,\cm-1 and 500\,\cm-1, they are dominated by Mo and Al lattice vibration modes.
In the Raman spectra shown in Fig.~\ref{Fig1_structure}c-e, we detect substantial mode intensity for the new Mo modes ($A^1$, $A^2$, and $A^3$) below 100\,\cm-1 and new C modes at around 500\,\cm-1 in the $XX$ scattering geometry, while those new $E$-symmetry modes involving the Al lattice vibration in-between 100\,\cm-1 and 350\,\cm-1 in the $X'Y'$ scattering geometry are 5 times weaker. 
This indicates that Mo and C atoms are mainly involved in the polar distortions below $T^*$. 
We note that previous $^{27}$Al NMR studies revealed the broadening of Al spectrum below $T^*$ and associated it to the gradual distortions near the Al sites toward a lower symmetric structure~\cite{Kuo_2012PhysRevB}. To resolve the lattice distortions below $T^*$, we have performed the single-crystal X-ray diffraction measurement of Mo$_3$Al$_2$C at 300\,K and 100\,K. We failed to detect the superlattice Bragg peaks in the diffraction patterns (Supplementary Note~X). 
Regardless of the origin for the polarization below $T^*$--proper or improper mechanism, substantial C-lattice contributions to the polar distortion, which are decoupled from the itinerant Mo $4d$ electronic states around the Fermi level, help to stabilize the polar metal phase of Mo$_3$Al$_2$C according to the decoupled electron mechanism~\cite{Benedek_2016_JMCC,Kim_2016_nature,Puggioni_2014_NC}
                                                                                                                                                                
     

\textbf{SC state.} Finally, we discuss the symmetry and the properties of the SC state in Mo$_3$Al$_2$C. The point group symmetry of Mo$_3$Al$_2$C in the SC state is $C_3$. It contains two irreducible representations $A$ and $E$. The $A$-symmetry SC order parameter preserves the three-fold rotational symmetry and the SC gap  could be isotropic or anisotropic.
The $E$-symmetry SC order parameter breaks the three-fold rotational symmetry and is a two-component order parameter. 
The lack of inversion symmetry in the low-crystallographic-symmetry Mo$_3$Al$_2$C necessarily leads to the mixing of singlet and triplet SC order parameters, thus creating an unconventional SC state.
Previous studies have reported that Mo$_3$Al$_2$C has a $s$-wave-like nodeless gap~\cite{Zhigadlo_2018PRB,Bauer_2010_PhysRevB,Bonalde_2011PhysRevB,Koyamado_2013_JPSJ}. However, a recent muon spin relaxation/rotation ($\mu$SR) study revealed that the ratio $T_{c}/\lambda_{\text{eff}}^{-2}$ ($\lambda_{\text{eff}}$ is the effective London penetration depth) is comparable to the class of unconventional superconductors in the Uemura plot, which points towards an unconventional pairing mechanism in the isostructural  compound W$_3$Al$_2$C~\cite{Gupta_2021PhysRevB}. The study of the relationship between the polar structure and the symmetry of the superconducting order parameters in Mo$_3$Al$_2$C is beyond the scope of this current work but will motivate future studies.


%
         
In summary, we identify Mo$_3$Al$_2$C as a new category of superconductor with structural polarity and chirality. Our results open up an avenue to discover more CDW-driven polarization in noncentrosymmetric metals as well as polar superconductors. 
Materials in this category could have new physical properties and applications, motivating future studies of switchable ferroelectric superconductivity~\cite{Jindal_2023Nature}, nonreciprocal charge transport~\cite{Tokura_2018NC,Nadeem_2023review}, and unconventional paring mechanism~\cite{Yip_2014Review,Kallin_2016}. \\    
     

\textbf{Methods}

\textbf{Single crystal preparation and characterization.}
Mo$_3$Al$_2$C single crystals were grown using a slow cooling method in a sealed alumina tube. Powders of Mo : Al : C in molar ratio 3 : 4 : 1 were mixed and pre-reacted in a sealed vacuum quartz tube at 1000$^{\circ}$C for 5 hours. The product was filled into an one-end-closed alumina tube. Then, the open end was sealed in a laser floating zone furnace in an Argon flow. The sealed alumina tube was heated to 1650$^{\circ}$C, kept for 5 hours, 5 $^{\circ}$C/h cooled to 1250$^{\circ}$C, then 100 $^{\circ}$C/h cooled to room temperature. Millimeter size crystals are mechanically separated from the product, and the single-crystallinity is confirmed by X-ray Laue diffraction and polarized light optical microscope observation.                        

Electric transport measurements were carried out in a standard four-point probe method in the (1~0~0) plane in the He exchange gas environment using a PPMS on both cooling-down and warming-up processes.

Magnetic susceptibility measurements were carried out in a Quantum Design SQUID magnetometer in He exchange gas environment in zero-field cooled (ZFC) and field-cooled (FC) processes.

The crystal structure was determined using a Bruker single crystal x-ray diffractometer. The structure was refined using the SHELXTL Software Package. 
Additionally, we have performed Laue diffraction to orientate Mo$_3$Al$_2$C crystals before the sample polishing.  

\textbf{Polarization-resolved Raman spectroscopy.}
The nature exposed surface of Mo$_3$Al$_2$C is (1~1~0) crystallographic plane. To decompose the Raman signal into separate irreducible representations, we choose to work with the high symmetry (1~0~0) crystallographic plane. The as-grown sample is polished to expose its (1~0~0) crystallographic plane with a lapping film (1\,$\mu$m, Buehler). 
A strain-free area for Raman scattering measurement is examined by a Nomarski image. The strain-free area is further examined by comparing the phonon linewidth obtained on the as-grown (1~1~0) crystallographic plane and the polished (1~0~0) crystallographic plane. We did not find any noticeable polishing-induced linewidth broadening.                             
                                   
The polished crystals with (1~0~0) plane used for the Raman scattering study were positioned in a continuous helium flow optical cryostat. The Raman measurements
were mainly performed using the Kr$^+$ laser line at 647.1\,nm (1.92\,eV) in
a quasibackscattering geometry along the crystallographic $c$ axis.
The excitation laser beam was focused into a $50\times100$ $\mu$m$^2$
spot on the $ab$ plane, with the incident power around 12\,mW. 
For the measurement below $T_c$, an elongated laser spot ($50\times600$ $\mu$m$^2$) and a weak laser power 1-2\,mW is used to reduce laser heating.
The scattered light was collected and analyzed by a triple-stage Raman
spectrometer and recorded using a liquid nitrogen-cooled
charge-coupled detector. 
Linear and circular polarizations are used in this study to decompose the Raman data into different irreducible representations.
The instrumental resolution was maintained better than 1.5\,\cm-1. 
All linewidth data presented in this paper have been corrected for the instrumental resolution by fitting the Raman peaks using a Voigt profile.
The temperature shown in this paper has been corrected for laser heating (Supplementary Sec.~XI).

All spectra shown were corrected for the spectral response of the spectrometer and charge-coupled detector. The obtained Raman intensity $I
_{\mu v}$, which is related to the Raman response $\chi''(\omega,T)$: $I_{\mu v}(\omega, T)=[1+n(\omega, T)] \chi_{\mu \nu}^{\prime \prime}(\omega, T)$. Here $\mu (v)$ denotes the polarization of the incident (scattered) photon, $\omega$ is energy, $T$ is temperature, and $ n(\omega, T)$ is the Bose factor.
                    
The Raman spectra have been recorded from the $ab$ (1~0~0) surface for scattering geometries denoted as $\mu v = XX, XY, X'X', X'Y', RR, RL$, which is short for $Z(\mu v)Z$ in Porto's notation, where $X$ and $Y$ denotes linear polarization parallel to the crystallographic axis [1~0~0] and [0~1~0], respectively; $X' $ and $Y'$ denotes linear polarization parallel to [1~1~0] and [1~$\bar{1}$~0], respectively; $R=X+iY$ and $L=X-iY$ denote the right- and left-circular polarizations, respectively; The $Z$ direction corresponds to the $c$ axis perpendicular to the (1~0~0) plane. 

\textbf{Transmission electron microscopy measurement.}
Single crystal of Mo$_3$Al$_2$C were polished, followed by Ar-ion milling and studied using a JEOL-2010F field-emission transmission electron microscopy equipped cryogenic sample stage.

\textbf{Group-theoretical analysis.}
Group theoretical predictions were performed using the tool provided in the Isotropy Software Suite and the Bilbao Crystallographic Server~\cite{Bilbao_1, Bilbao_4, Hatch2003}.

The correlation table [Table~\ref{CorrelationTable}] lists all ten nontrivial subgroups of $O$ and provides the correspondences of irreducible representations. Each subgroup is generated by a subset of the symmetry elements of the parent group, as indicated in the table header. 
Cubic point groups have a complicated product structure involving a tetragonal (or orthorhombic) and a trigonal subsymmetry.
Notably, there are two distinct $D_2$ subgroups within $O$; one is associated with the tetragonal subsymmetry, while the other combines symmetry elements from both classes. Additionally, the table includes a redundant $D_4^*$ group, which, although identical to the standard-oriented $D_4$, is constructed from different generators; the only difference between the two entries lies in the $B_1$ and $B_2$ labels. Based on Table~\ref{CorrelationTable}, only the $C_3$ and $D_3$ subgroups show that $E$ irreducible representation does not split while $T_2$ irreducible representation splits into two irreducible representations.

   
\begin{table}[!t]
\caption{\label{CorrelationTable} Correlation table for 
point group $O$.}   
\begin{tabular}{|c|c|c|c|c|c|c|c|c|c|c|}
\hline $O$ & \begin{tabular}{l}
\begin{tabular}{c}
$T$ \\
$c_3, c_2$
\end{tabular}
\end{tabular} & \begin{tabular}{l}
\begin{tabular}{c}
$D_4$ \\
$c_4, c_2$
\end{tabular}
\end{tabular} & \begin{tabular}{l}
\begin{tabular}{c}
$D_4^*$ \\
$c_4, c_2^{\prime}$
\end{tabular}
\end{tabular} & \begin{tabular}{l}
\begin{tabular}{l}
$C_4$ \\
$c_4$
\end{tabular}
\end{tabular} & \begin{tabular}{l}
\begin{tabular}{c}
$D_2$ \\
$c_2, c_2$
\end{tabular}
\end{tabular} & \begin{tabular}{l}
\begin{tabular}{c}
$D_2$ \\
$c_2, c_2^{\prime}$
\end{tabular}
\end{tabular} & \begin{tabular}{l}
\begin{tabular}{l}
$C_2$ \\
$c_2$
\end{tabular}
\end{tabular} & \begin{tabular}{l}
\begin{tabular}{c}
$D_3$ \\
$c_3, c_2^{\prime}$
\end{tabular}
\end{tabular} & \begin{tabular}{l}
\begin{tabular}{l}
$C_3$ \\
$c_3$
\end{tabular}
\end{tabular} & \begin{tabular}{l}
\begin{tabular}{l}
$C_2$ \\
$c_2^{\prime}$
\end{tabular}
\end{tabular} \\
\hline$A_1$ & $A$ & $A_1$ & $A_1$ & $A$ & $A$ & $A$ & $A$ & $A_1$ & $A$ &$ A$ \\
\hline$A_2$ & $A$ & $B_1$ & $B_2$ & $B$ & $A$ & $B_1$ & $A$ & $A_2$ & $A$ & $B$ \\
\hline $E$ & $E$ & $A_1+B_1$ & $A_1+B_2$ & $A+B$ & $2 A$ & $A+B_1$ & $2 A$ & $E$ & $E$ & $A+B$ \\
\hline$T_1$ & $T$ & $A_2+E$ & $A_2+E$ & $A+E$ & $B_1+B_2+B_3$ & $B_1+B_2+B_3$ & $A+2B$ & $A_2+E$ & $A+E$ & $A+2 B$ \\
\hline$T_2$ & $T$ & $B_2+E$ & $B_1+E$ & $B+E$ & $B_1+B_2+B_3$ & $A+B_2+B_3$ & $A+2B$ & $A_1+E$ & $A+E$ & $2 A+B$ \\
\hline
\end{tabular}   
\end{table}  

\textbf{First-principles calculations.} 
First-principles calculations were performed using the \textsc{vasp} code which uses projector-augmented wave formalism~\cite{Kresse199607,Kresse1996_PhysRevB,Blochl1994,Kresse1999}. DFT+U formalism was not used due to the metallic nature of Mo$_3$Al$_2$C with only $4d$ transition metal cations. Similarly, spin-orbit coupling was ignored. The PBEsol exchange correlation functional~\cite{PBEsol} was used due to the success of this generalized gradient approximation in reproducing the lattice parameters of solids, which is particularly important when studying lattice instabilities. The direct method, which involves displacing atoms one by one and using the Hellmann-Feynman forces to obtain force constants, was employed to calculate the phonon frequencies. A Gaussian smearing of 200\,meV for the electronic occupations was employed, along with a $8\times8\times8$ k-grid to approximate the Brillouin zone integrals. While the unstable phonon frequencies exhibit some sensitivity to these parameters, which is often the case for charge-density wave instabilities~\cite{Christensen_arxiv2021}, our calculations with lower values of smearing or denser k-grids did not give qualitatively distinct results.

\textbf{Data availability}\\ 
The transport and Raman data generated in this study have been deposited in the figshare database under accession code https://doi.org/10.6084/m9.figshare.27192603. More data are available from the corresponding authors upon request.


\vspace{1cm}
\textbf{Acknowledgments} \\
The spectroscopic work conducted at Rutgers 
(S.F.W. and G.B.) was supported by the NSF Grant No.~DMR-2105001. 
The sample growth, characterization, and TEM work (X.H.X. F.T.H and S.W.C.) were supported by the DOE under Grant No.~DOE: DE-FG02-07ER46382.
The theoretical work conducted at the University of Minnesota (E. R. and T. B.) was supported by the NSF CAREER Grant No.~DMR-2046020.
The work at NICPB was supported by the European Research Council (ERC) under the European Union’s Horizon 2020 research and innovation programme grant agreement No.~885413.

\textbf{Author Contributions} \\
G.B. and S.W.C designed and supervised the experiments. S.F.W and G.B.
acquired and analyzed the Raman data. X.H.X. and S.W.C synthesized the single crystal.
F.T.H and S.W.C. acquired and analyzed the TEM data. E.R and T.B. did the supercell, domain, and free-energy model analysis and first-principles phonon calculations.
All authors contributed to the discussion and writing of the manuscript.

\textbf{Competing interests} \\
The authors declare no competing interests.

\textbf{Materials \& Correspondence}\\
Correspondence and requests for materials should be addressed to Shangfei~Wu, Turan~Birol, Sang-Wook~Cheong, and Girsh~Blumberg.

%
\end{document}